\documentclass[twocolumn, pra, 10pt, floatfix]{revtex4-2}
\usepackage{graphicx}
\usepackage[justification=raggedright, font=footnotesize]{caption}
\usepackage{float}
\usepackage{upgreek}

\begin{document}

% Title / authors / abstract
    \title{Double quantum dots with quenched charging energy in PbTe nanowires}
    \author{Seth Byard\textsuperscript{1}}
    \author{Maksim Gomanko\textsuperscript{1}}
    \author{Adam Raynolds\textsuperscript{1}}
    \author{Susheng~Tan\textsuperscript{2}}
    \author{Tongxie Zhang\textsuperscript{3}}
    \author{Shixiong Zhang\textsuperscript{3}}  
    \author{Sergey M. Frolov\textsuperscript{1}}\email{Corresponding author: frolovsm@pitt.edu}
    % Write all affiliations together for formatting reasons
    \affiliation{\looseness=-1
    \textsuperscript{1}Department of Physics \& Astronomy, University of Pittsburgh, Pittsburgh, PA 15260, USA
    \\
    \textsuperscript{2}Department of Electrical \& Computer Engineering and Petersen Institute of Nanoscience \& Engineering, University of Pittsburgh, Pittsburgh, PA 15260, USA
    \\
    \textsuperscript{3}Department of Physics, Indiana University Bloomington, Bloomington, IN 47405, USA
    }

\begin{abstract}
    We investigate double quantum dots defined by electrostatic gating in semiconductor PbTe nanowire devices. We perform transport measurements to obtain charge stability diagrams distinguished by negligible separation between paired triple points and by the spin degeneracy of all transport resonances at zero magnetic field. We show a fourfold splitting of high-bias stability diagram triangles in an applied magnetic field to illustrate the lifting of this spin degeneracy. We also identify patterns of narrow transport resonances in these high-bias triangles and discuss their possible physical origins. Our results represent a step towards the realization of PbTe-based spin qubits.
\end{abstract}
    \maketitle
    
% Main text
    %
%
\section*{Introduction}

Semiconductor spin qubits are an attractive quantum computing platform due to factors such as their scalability~\cite{Langrock, Neyens, Maurand, George} and long achievable coherence times~\cite{Bluhm, Veldhorst, Kobayashi}. Contemporary spin qubit research is largely focused on group IV semiconductors and continues to see many successes with these~\cite{Mills, Hendrickx, Xue}. However, the development of a practical quantum computer will still require tremendous scientific and technological advances~\cite{Wecker}, and we argue that investigating materials is an important part of this effort~\cite{deLeon}.

PbTe is a group IV-VI semiconductor characterized by its rocksalt crystal structure, with 4 band minima at L points in the Brillouin zone, and a narrow direct band gap of $0.19\:\mathrm{eV}$ at room temperature which reduces at low temperatures. PbTe has low effective electron mass, allowing for higher carrier mobility~\cite{Dzundza}. It also features large spin-orbit interaction and g-factor~\cite{Peres, Gomanko, tenKate}, which may be useful for all-electrical control of quantum dot spins~\cite{Golovach} at low magnetic fields. Both Pb and Te have stable isotopes with zero nuclear spin, suggesting that PbTe could in principle be isotopically enriched to minimize dephasing due to hyperfine interaction~\cite{Tyryshkin}. These properties make PbTe a promising potential host for spin qubits~\cite{NadjPerge, Kloeffel, Huang}. 

An interesting feature of PbTe is its large static dielectric constant that exceeds $1000$ at cryogenic temperatures~\cite{Springholz}. This is expected to screen charge impurities and through this increase the mean free path~\cite{Springholz}. Another consequence of an extremely large dielectric constant is the dramatically reduced charging energy in quantum dots. Single quantum dots with quenched Coulomb blockade have been realized~\cite{Gomanko, tenKate, Cheng_Levy}. However, in the context of spin qubits it is important to understand how these phenomena manifest in double quantum dots~\cite{Burkard, Nichol, Xue}.

In this work, we investigate double quantum dots (DQDs) defined by electrostatic gating in PbTe nanowires. We perform transport measurements to obtain charge stability diagrams which have doubled occupancy for each dot at zero magnetic field due to quenched charging energy. In combination with vanishing interdot capacitance, this manifests in overlapping charge degeneracy points as opposed to the pairs of distinct triple points typical for DQDs with Coulomb blockade. We demonstrate the lifting of this spin degeneracy in an externally applied magnetic field, where four charge degeneracy points split out of a single point at zero field. Lastly, we find narrow transport resonances within high-bias stability diagram triangles which may signify interdot transitions. These findings inform future efforts to demonstrate coherent control of spins in PbTe and eventually evaluate spin coherence times.
    \section*{Methods}
The PbTe nanowires used in this study are grown using chemical vapor deposition (CVD) via a Au-catalyzed vapor-liquid-solid (VLS) method, following the growth conditions described in Ref.~\cite{ZhangTongxie}. The developed process makes use of an increased volume of Te precursor during growth, which is found to greatly improve results due to Te facilitating the precipitation of Pb from the alloy particles. The nanowires’ dimensions can vary significantly, but nanowires of length 5–10 µm and width 100–200 nm, ideal for device fabrication, are still sufficiently abundant and easy to identify. We assess the crystallinity and cubic structure of our PbTe nanowires through transmission electron microscopy (TEM) and associated techniques (Figure~\ref{fig:supp_tem} in the supplementary information).
\par
Figure~\ref{fig:device}(a)-(b) presents a typical device from this study. Devices are fabricated on $\mathrm{Si/SiO_2}$ substrates patterned with arrays of bottom gates covered in thin dielectric HfO\textsubscript{2}. The arrays consist of sets of 5 narrow gate electrodes with wider metal pads lying in-between each set. Nanowires are transferred onto the arrays, then metallic source/drain contacts are added such that a section of each nanowire is left open over a set of gates. This deposition is done at a series of 3 angles ($0^\circ$, $\pm 45^\circ$), as shown in Figure~\ref{fig:device}(c), for improved contact. The device is coated in thin HfO\textsubscript{2}, then a large top gate electrode is added. A more detailed description of the fabrication process may be found in the supplementary information. 

We electrostatically define DQDs in nanowires by applying voltages to top and bottom gates. The primary device of this study has one bottom gate accidentally connected to an adjacent metal pad. This greatly amplifies the gate's effect, but also makes it further difficult to predict the location and geometry of defined dots. Measurements are done in a dilution refrigerator equipped with a vector magnet at a base temperature of $\sim 50$ mK. 

\begin{figure}[tb]
    \includegraphics[width=\columnwidth]{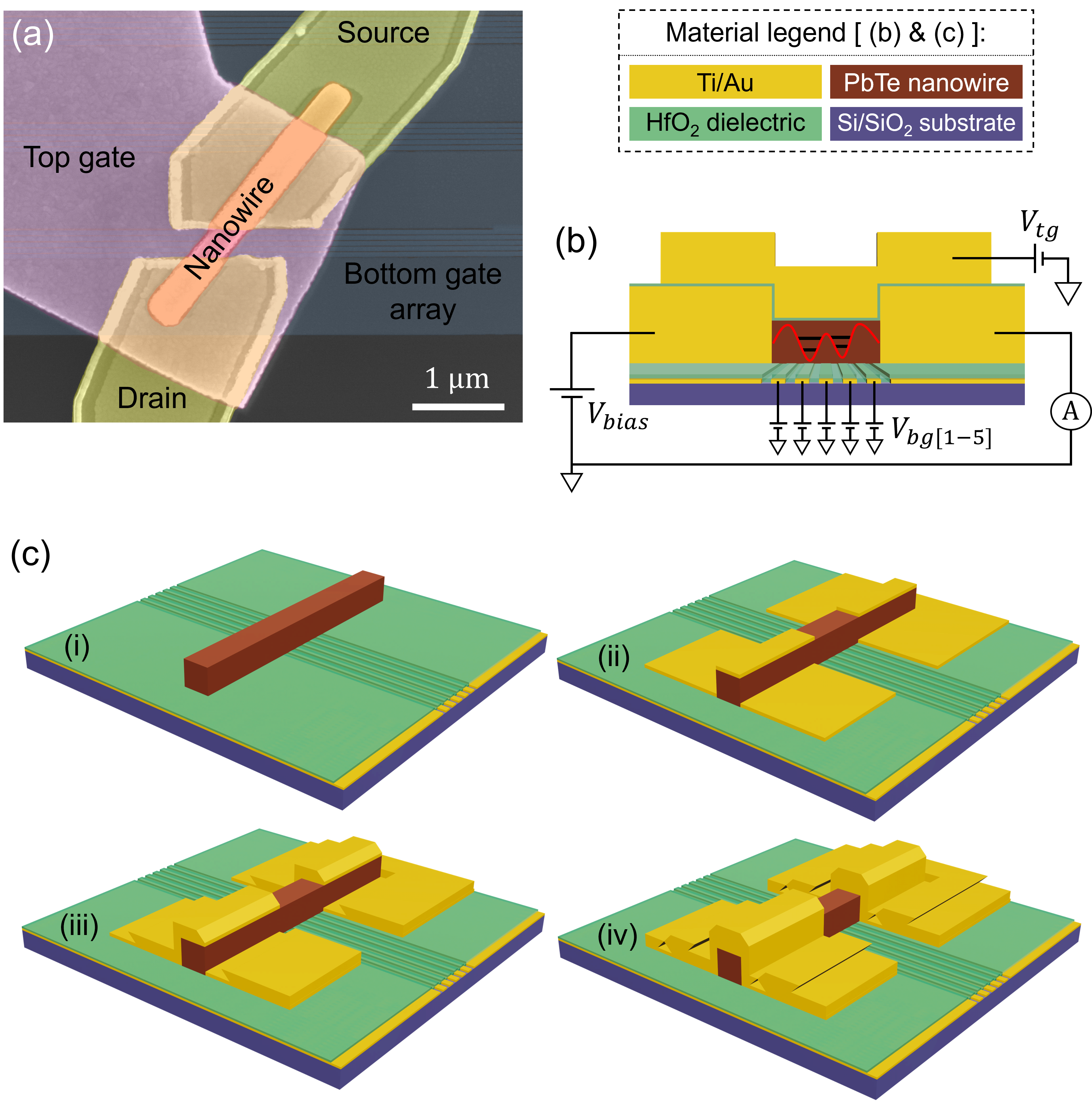}
    \caption{(a) False-color SEM image of the primary device used in this study. (b) Simplified side-view schematic of a device and of the measurement setup. (c) 3D schematic of an in-progress device during the angled contact deposition process, with a cross section exposed for clarity. Dimensions of shown features are to scale. (i) Bare nanowire over a gate array. (ii) Direct top-down deposition. (iii) $+45^\circ$ deposition. (iv) $-45^\circ$ deposition.}
    \label{fig:device}
\end{figure}

We observe that many devices undergo dramatic changes in contact resistance over a period of time. We generally measure initial resistances in the range of $\sim$100~k$\mathrm{\Omega}$--1~M$\mathrm{\Omega}$, but after roughly 1 week or more passes, we find values of $\sim$1--10 k$\mathrm{\Omega}$. All devices examined in this work have undergone this process. We note that devices go through such a change only after the second HfO\textsubscript{2} layer is deposited, so we argue that the HfO\textsubscript{2} itself must be relevant. For example, Hf atoms may diffuse into the nanowire's native oxide and yield doping~\cite{Nagarajan}.
\par
    \section*{Results}
\begin{figure}[tb]
    \includegraphics[width=\columnwidth]{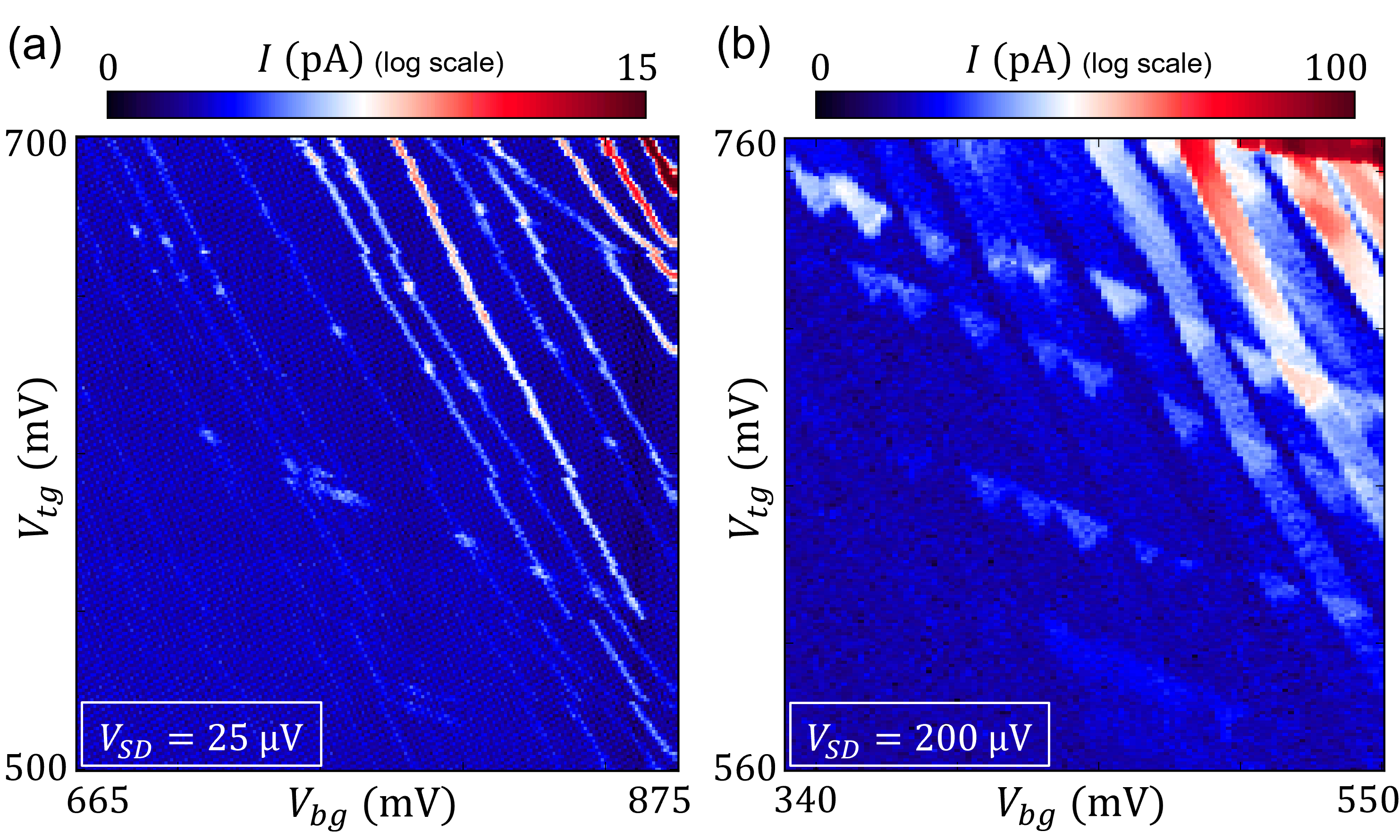}
    \caption{DQD charge stability diagrams at low [(a)] and high [(b)] bias. No magnetic field is applied. We used different sets of gate potentials for the two scans since the spectrum had shifted due to an uncontrolled charge jump. Additional bottom gates ``bg3'' and ``bg4'' are set to $V_{bg3}=V_{bg4}=-250$~mV [(a)] and $V_{bg3}=V_{bg4}=+700$~mV [(b)]. A linecut is subtracted from all rows of the left plot to remove a background signal.}
    \label{fig:data_wide}
\end{figure}

Figure~\ref{fig:data_wide} shows experimental charge stability diagrams, representing current $I$ vs top and bottom gate potentials $V_{tg}$ and $V_{bg}$, at low and high source-drain biases. We observe two distinct sets of parallel lines which intersect to form bright spots of increased current, although many of the lines here are faint except at these intersections. The small, bright points sitting at the intersections at low bias evolve into triangles as the bias is increased. These behaviors are classic characteristics of a double quantum dot (DQD)~\cite{VanDerWiel, Hanson}. Each set of parallel resonances corresponds primarily to one of the dots, with the change in charge occupancy corresponding to transitions across the resonances. The bright points and bias triangles represent states of resonant transport across both dots due to the alignment of their electrochemical potential levels within the bias window. Significant transport is also seen here to occur along some resonances rather than solely at their intersections, likely due to co-tunneling processes~\cite{VanDerWiel}. Several abrupt shifts in the pattern are present here which tend to reoccur at similar values if a scan is repeated. We observe this behavior in other similar devices and attribute it to nearby charge traps, which may effectively act as uncontrolled dots coupled in parallel to a main dot~\cite{Byard}. The main dots' electron occupancies are unknown since the lowest filled levels cannot be clearly identified, but we estimate them to be on the order of between few to low tens of electrons per dot.
\par
We examine individual bias triangles to extract the gate lever arms, which are conversion factors between the voltage applied to a gate and the electrochemical potential of a dot. See Supplementary Figure~\ref{fig:supp_lever_arms} for details of this analysis. The calculated lever arms indicate that the bottom gate is most strongly coupled to the first dot, and the top gate to the second dot, where the ordering of dots is defined relative to the source/drain. However, the two lever arms of each gate differ from each other by less than a factor of 2, indicating that the capacitance or ``cross-talk'' between gates is significant.

Typical DQDs have pairs of distinct ``triple points'' near the resonances' intersections. The separation between paired points results in characteristic honeycomb patterns of hexagonal cells. The mutual capacitance $C_m$ between the two dots may then be found by measuring this separation~\cite{Hanson}. In contrast, our data show patterns of repeated parallelograms with individual points rather than pairs. We also observe only individual triangles at high bias rather than pairs. We understand these behaviors to be evidence of a limiting case where the spacing between paired triple points is vanishingly small such that they effectively overlap. The resulting charge degeneracy points then essentially lie between 4 charge regions instead of 3. Since the spacing between points is proportional to $C_m$, we conclude that these two dots are effectively capacitively uncoupled, i.e. $C_m\approx0$. This may be due to a combination of the screening effects of gates and PbTe's large dielectric constant.
\par
\begin{figure}[tb]
    \includegraphics[width=\columnwidth]{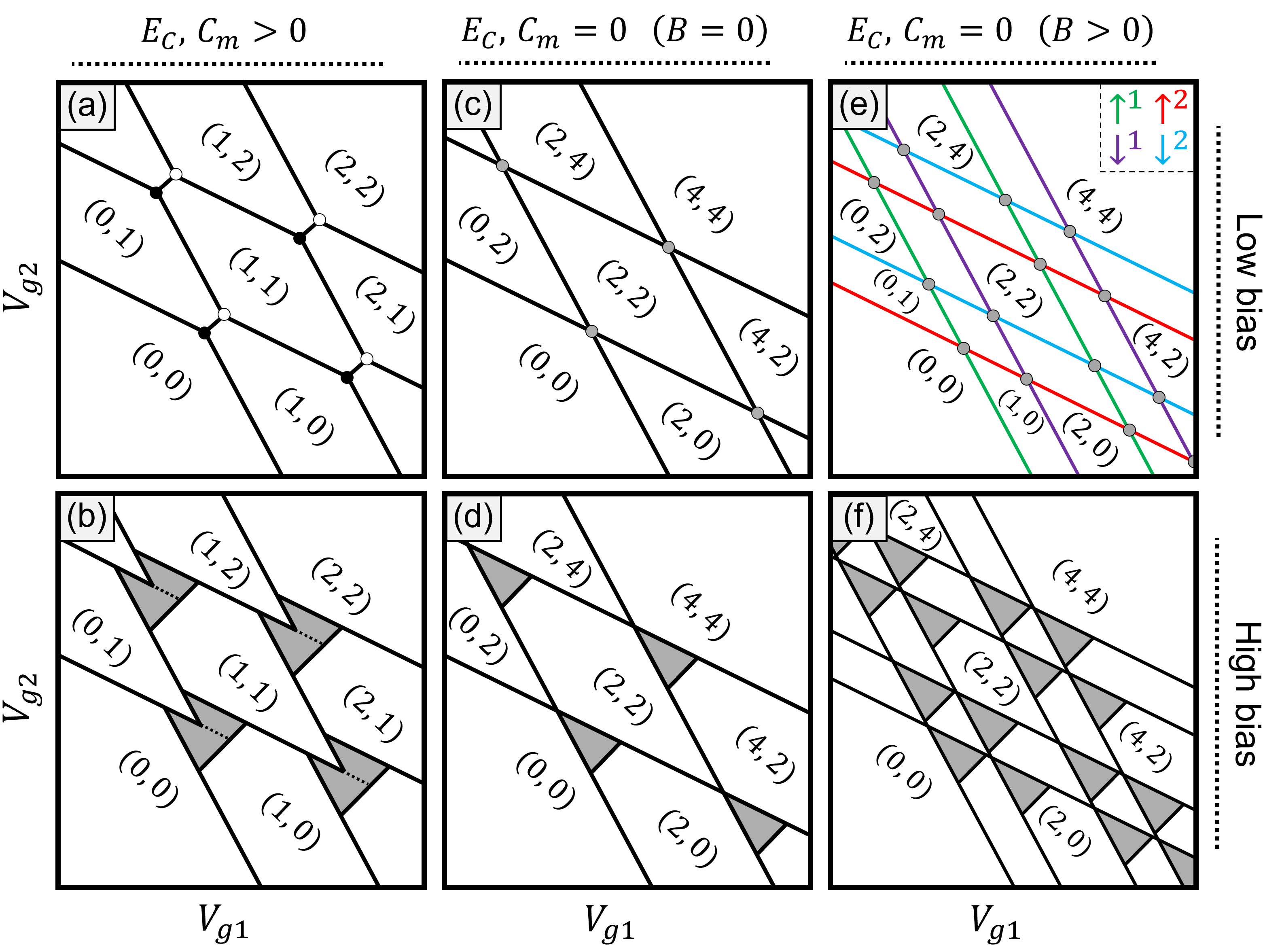}
    \caption{(a)-(f) Schematic charge stability diagrams, at low or high bias, for DQDs with finite or zero charging energy $E_C$ and mutual capacitance $C_m$. The case of $E_C,C_m=0$ is considered both without and with a magnetic field $\vec B$ applied. Avoided crossings due to interdot tunnel coupling are not shown. (e) Resonances' spin states ($\uparrow$ and $\downarrow$) and associated dots (1 or 2) are indicated as shown in the inset legend.}
    \label{fig:diagram_schematic}
\end{figure}

Quantum dots in PbTe have quenched charging energy $E_C$ due to the large dielectric constant~\cite{Gomanko, tenKate}. When both $E_C$ and $C_m$ are negligibly small, the resulting DQD stability diagrams show major differences from those of typical systems. Figure~\ref{fig:diagram_schematic} depicts a series of schematic diagrams which highlight these differences. We refer to refs.~\cite{VanDerWiel} and~\cite{Hanson} for details on typical DQD behaviors and stability diagrams.

When $E_C=0$, the dots' electrochemical potential levels are all spin-degenerate in the absence of an applied magnetic field, meaning that electrons are only loaded or unloaded in pairs from either dot. Applying a magnetic field $\vec B$ lifts this degeneracy via the Zeeman effect, resulting in a twofold splitting of every transport resonance and consequently in a fourfold splitting of every charge degeneracy point or bias triangle. Figure~\ref{fig:diagram_schematic}(e) illustrates this effect and indicates the association between each resonance and a spin state of one dot. Note that $\vec B$ does not noticeably impact $C_m$, meaning that each charge degeneracy point or bias triangle should still be viewed as an overlapping pair.
\par
\begin{figure}[tb]
    \includegraphics[width=0.97\columnwidth]{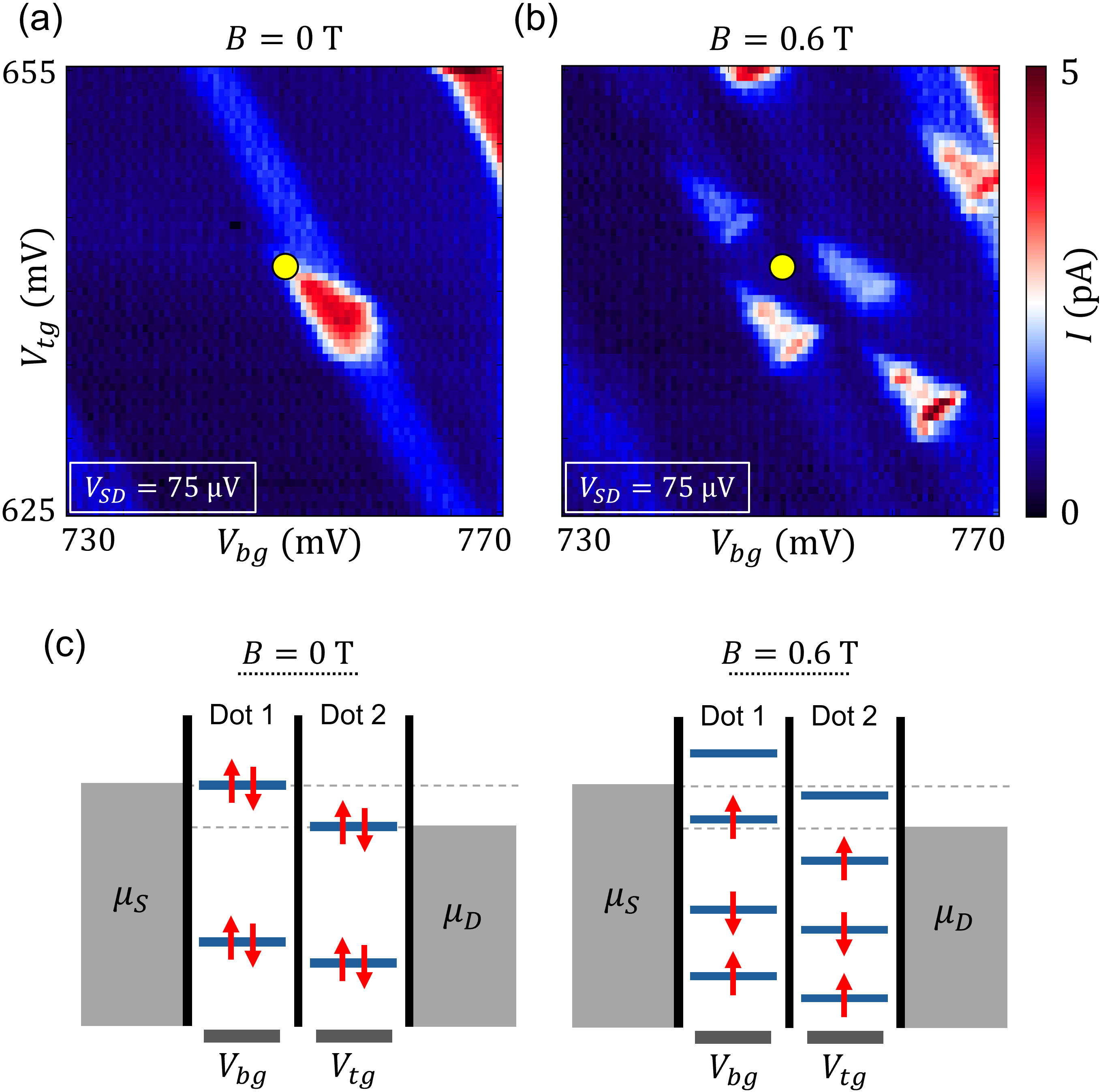}
    \caption{(a)-(b) Charge stability diagrams at a mid-range bias both without [(a)] and with [(b)] a magnetic field $\vec{B}$ applied. Additional bottom gates ``bg3'' and ``bg4'' are set to $V_{bg3}=V_{bg4}=-250$~mV. (c) Simplified schematic model of the DQD electrochemical potential levels roughly corresponding to the fixed set of gate voltages indicated by the yellow dots in (a)-(b), without and with $\vec{B}$ applied. This particular set of gate voltages is chosen semi-arbitrarily just for illustrative purposes.}
    \label{fig:data_degenerate}
\end{figure}

% Note: the paragraph below is placed in this block for formatting reasons. Its content probably belongs in the Fig 3 block.
We assume that the valley degeneracy is lifted. Otherwise, each transition can be additionally 4-fold degenerate. This assumption is supported by the fact that we do not observe a valley filling pattern~\cite{Sapmaz} in a different device (Device C) from this study in which we do observe Coulomb blockade. However, that device is based on an irregularly shaped nanowire which could facilitate intervalley coupling. Results from Device C are further discussed later in the main work and in Supplementary Figure~\ref{fig:supp_Cdata}. We also note a lack of evidence for valley degeneracy in previous work~\cite{Gomanko}, although the nanowires and devices differed from those used here. 

Figure~\ref{fig:data_degenerate}(a) presents experimental stability diagrams showcasing the fourfold splitting of a bias triangle in an applied magnetic field. This behavior is examined further in Figure~\ref{fig:data_degenerate}(b) through schematic diagrams of the dots' electrochemical potential levels for a representative set of gate voltages.
\par
\begin{figure}[tb]
    \includegraphics[width=\columnwidth]{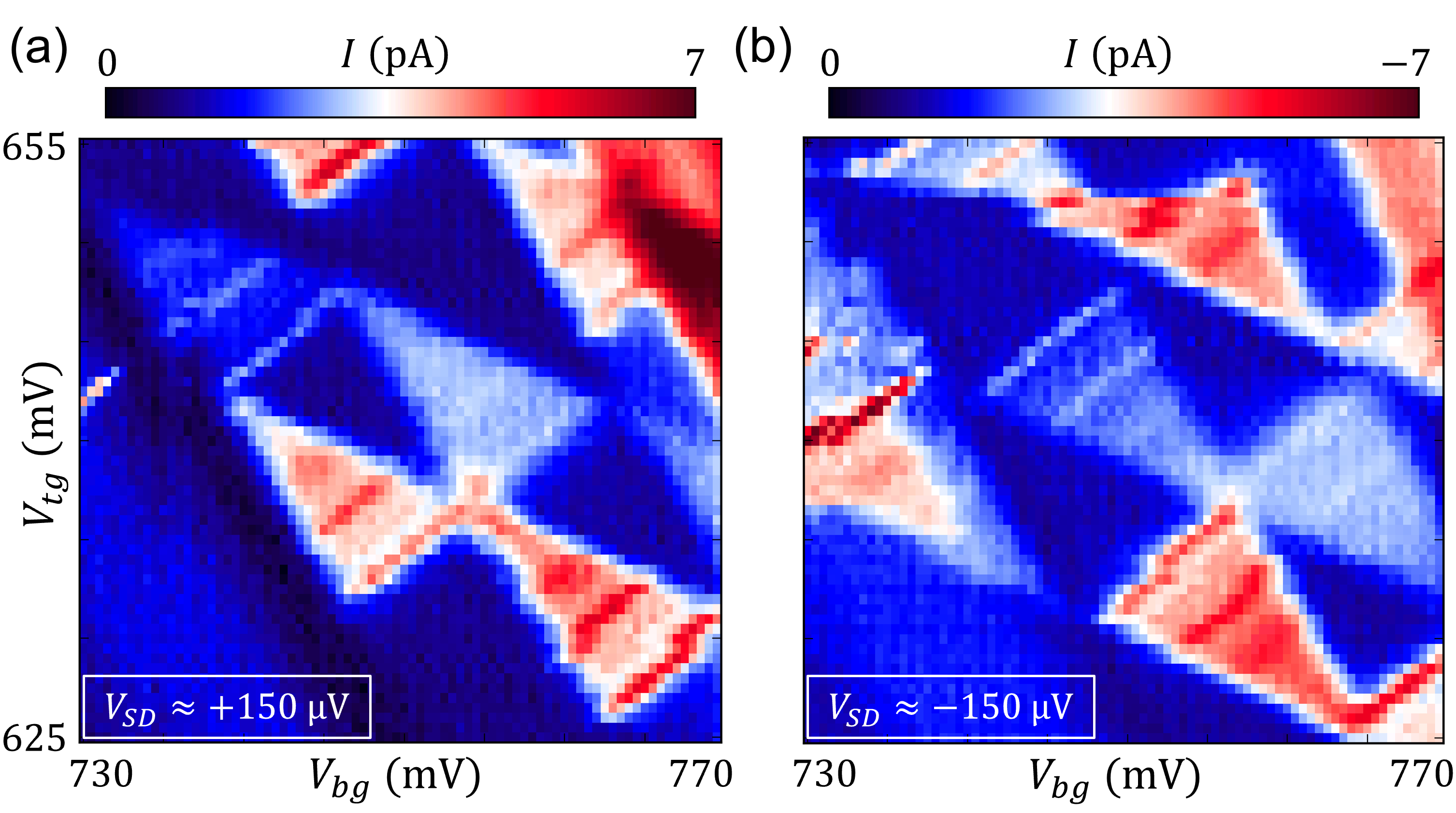}
    \caption{Charge stability diagrams at high positive and negative biases with magnetic field $B=0.8\:\mathrm{T}$ applied. Bottom gates ``bg3'' and ``bg4'' are set to ${V_{bg3}=V_{bg4}=-250}$~mV. We attribute the triangles’ slight size variation between scans to a bias offset of roughly $-10\:\mathrm{\upmu V}$. Sharp interruptions in the signal may be due to coupled charge traps~\cite{Byard}.}
    \label{fig:data_excited}
\end{figure}

Close inspection of the data in Figure~\ref{fig:data_degenerate} reveals the presence of narrow lines that run parallel to the triangle bases. These additional resonances are most apparent at finite magnetic field, but are also present at zero field. By increasing the source-drain bias and magnetic field, we may examine these features in more detail as shown in Figure~\ref{fig:data_excited}. Up to 3 clear resonances per triangle can be seen, spaced by $\sim$30--60 $\upmu$eV. Other examples of features within the high-bias triangles are presented in the supplementary information (Figures \ref{fig:supp_Cdata}, \ref{fig:supp_Ddata}).

Similar features reported in dots with significant charging energy are typically attributed to resonant transport through higher energy ``excited'' DQD states~\cite{VanDerWiel, Hanson}. However, in the absence of Coulomb blockade any such state should be visible at zero bias and generate its own charge degeneracy point. Similarly, discrete states defined in the device's leads, which could be considered additional quantum dots in series, would generate an independent set of charge stability regions visible even at zero bias, though in principle their visibility could be enhanced inside the bias triangles due to the increased signal there. Excited states originating from a fine spin structure~\cite{Hanson} seem unlikely due to the fact that the resonances and their spacings appear similar whether in zero or large magnetic field. Resonances similar to those in Figures \ref{fig:data_degenerate} and \ref{fig:data_excited} have been reported in quantum dots due to either photon-assisted transport~\cite{Mavalankar} or interactions with phonons~\cite{Hartke}. We anticipate that further work can help elucidate the nature of these bias triangle resonances.
\par
Along with the primary device discussed in this work, we report results from our measurements of the additional Devices B and C. Data from these, and from a fourth Device D, are presented in the supplementary information along with more detailed descriptions.

Device B shows a particularly strong response to the bottom gate arrays. Through fine-tuning of the gate potentials, we can freely shift between distinct transport regimes dominated by single-dot or multi-dot behavior, as evidenced by the charge stability diagrams showing roughly parallel resonances that evolve into a honeycomb-like pattern. The data appear to indicate the presence of additional interacting dots, such that a pure DQD regime is difficult to achieve except within certain small gate ranges. We find from examining resonances' avoided crossings that there is a notable tunnel coupling between the interacting dots. The spacing also suggests a mutual capacitance $C_m$ between dots that is non-negligible, although still small.

Device C is atypical among our PbTe devices in that it has a clear non-zero charging energy $E_C$. We thus observe effects such as Coulomb blockade that are normally not present. We also identify the presence of interdot tunnel coupling and mutual capacitance $C_m$. We note that Device C is fabricated with a nanowire that has some clear deformation, so we argue that the atypical properties are likely due to this.
\par
\section*{Conclusions}

We define and characterize double quantum dots in PbTe nanowire devices through the use of electrostatic gating. We obtain charge stability diagrams with each dot doubly-occupied at zero magnetic field due to spin-degeneracy.  We study the resulting charge degeneracy points which split fourfold in an externally applied magnetic field due to the lifting of spin degeneracy within each dot. We further show that high-bias triangles have an internal structure of multiple distinct transport resonances which do not appear within the stability diagrams near zero bias. 

An important open question is whether spin-selective transport phenomena can occur in the absence of Coulomb blockade. At zero external magnetic field, each orbital energy level in PbTe quantum dots is occupied by both spin-up and spin-down electrons, making the most basic Pauli spin blockade not feasible~\cite{Ono, Johnson}. From the spin qubit point of view, this presents a limitation since spin blockade is often used to initialize and read out quantum dot spin states~\cite{Petta, Koppens}. An alternative is to use the energy readout facilitated by charge sensing~\cite{Elzerman, Petersson, Crippa}. Another interesting direction of future study is whether the large dielectric constant can in fact facilitate charge noise screening~\cite{Springholz} and increase spin qubit coherence times that may be affected by charge noise through spin-orbit interaction. The unconventional characteristics of PbTe continue to make it an interesting material to study quantum dot phenomena. 
\section*{Duration \& volume of study}

This study was primarily done between early 2024 and early 2025. During this time, the devices' design and fabrication were developed, and then multiple sets of measurements were carried out for several months in total. Measurements focused on DQDs were taken for 4 devices and yielded over 5000 datasets. Of these,  $\sim$2100 datasets are from the primary device examined in this work.
\section*{Data availability}

Data are available through Zenodo (DOI: 10.5281/zenodo.17031945). This includes nearly all data obtained from the 4 major devices of this study. The only data excluded is that which was collected as part of a separate but parallel study involving one of these same devices. The repository also includes copies of all figures and individual TEM and SEM images used in this work, including in the supplementary information.
\section*{Summary of supplementary information}

An expanded discussion of the device fabrication process described in the main work is provided. Figure~\ref{fig:supp_growthchip} shows a representative SEM image of a section of a nanowire growth sample, which showcases the notable variation in nanowire size. Figure~\ref{fig:supp_tem} shows representative images from our TEM analysis of nanowires, including HR-TEM scans and diffraction patterns. Figure~\ref{fig:supp_more_devices} shows SEM images of the 4 devices studied for this work: Devices A, B, C, and D. Figure~\ref{fig:supp_lever_arms} provides details on how we calculate gate lever arms from analysis of bias triangles. The key parameters are visually indicated on an experimental stability diagram, and extracted values are listed. Figure~\ref{fig:supp_extraAdata} presents additional stability diagrams from Device A that illustrate the fourfold splitting of bias triangles in a magnetic field at a wide gate range. Figure~\ref{fig:supp_Bdata} presents representative data from Device B. These data show a high degree of gate control for a multi-dot system, including control of interdot tunnel coupling. Figure~\ref{fig:supp_Cdata} presents representative data for Device C. These data show features with finite charging energy, which we attribute to the atypical physical features of the nanowire used. Finally, Figure~\ref{fig:supp_Ddata} presents representative data for Device D. This device yielded very noisy and unstable results, but the data still show certain key features observed in other PbTe DQDs.
\section*{Acknowledgments}

Device fabrication and measurement work is supported by the U.S. Department of Energy Basic Energy Sciences under grant DE-SC-0019274.
    
% Bibliography
    \bibliography{mainBib}

%apsrev4-2.bst 2019-01-14 (MD) hand-edited version of apsrev4-1.bst
%Control: key (0)
%Control: author (8) initials jnrlst
%Control: editor formatted (1) identically to author
%Control: production of article title (0) allowed
%Control: page (0) single
%Control: year (1) truncated
%Control: production of eprint (0) enabled
\begin{thebibliography}{40}%
\makeatletter
\providecommand \@ifxundefined [1]{%
 \@ifx{#1\undefined}
}%
\providecommand \@ifnum [1]{%
 \ifnum #1\expandafter \@firstoftwo
 \else \expandafter \@secondoftwo
 \fi
}%
\providecommand \@ifx [1]{%
 \ifx #1\expandafter \@firstoftwo
 \else \expandafter \@secondoftwo
 \fi
}%
\providecommand \natexlab [1]{#1}%
\providecommand \enquote  [1]{``#1''}%
\providecommand \bibnamefont  [1]{#1}%
\providecommand \bibfnamefont [1]{#1}%
\providecommand \citenamefont [1]{#1}%
\providecommand \href@noop [0]{\@secondoftwo}%
\providecommand \href [0]{\begingroup \@sanitize@url \@href}%
\providecommand \@href[1]{\@@startlink{#1}\@@href}%
\providecommand \@@href[1]{\endgroup#1\@@endlink}%
\providecommand \@sanitize@url [0]{\catcode `\\12\catcode `\$12\catcode `\&12\catcode `\#12\catcode `\^12\catcode `\_12\catcode `\%12\relax}%
\providecommand \@@startlink[1]{}%
\providecommand \@@endlink[0]{}%
\providecommand \url  [0]{\begingroup\@sanitize@url \@url }%
\providecommand \@url [1]{\endgroup\@href {#1}{\urlprefix }}%
\providecommand \urlprefix  [0]{URL }%
\providecommand \Eprint [0]{\href }%
\providecommand \doibase [0]{https://doi.org/}%
\providecommand \selectlanguage [0]{\@gobble}%
\providecommand \bibinfo  [0]{\@secondoftwo}%
\providecommand \bibfield  [0]{\@secondoftwo}%
\providecommand \translation [1]{[#1]}%
\providecommand \BibitemOpen [0]{}%
\providecommand \bibitemStop [0]{}%
\providecommand \bibitemNoStop [0]{.\EOS\space}%
\providecommand \EOS [0]{\spacefactor3000\relax}%
\providecommand \BibitemShut  [1]{\csname bibitem#1\endcsname}%
\let\auto@bib@innerbib\@empty
%</preamble>
\bibitem [{\citenamefont {Langrock}\ \emph {et~al.}(2023)\citenamefont {Langrock} \emph {et~al.}}]{Langrock}%
  \BibitemOpen
  \bibfield  {author} {\bibinfo {author} {\bibfnamefont {V.}~\bibnamefont {Langrock}} \emph {et~al.},\ }\href@noop {} {\bibfield  {journal} {\bibinfo  {journal} {PRX Quantum}\ }\textbf {\bibinfo {volume} {4}},\ \bibinfo {pages} {020305} (\bibinfo {year} {2023})}\BibitemShut {NoStop}%
\bibitem [{\citenamefont {Neyens}\ \emph {et~al.}(2024)\citenamefont {Neyens} \emph {et~al.}}]{Neyens}%
  \BibitemOpen
  \bibfield  {author} {\bibinfo {author} {\bibfnamefont {S.}~\bibnamefont {Neyens}} \emph {et~al.},\ }\href@noop {} {\bibfield  {journal} {\bibinfo  {journal} {Nature}\ }\textbf {\bibinfo {volume} {629}},\ \bibinfo {pages} {80} (\bibinfo {year} {2024})}\BibitemShut {NoStop}%
\bibitem [{\citenamefont {Maurand}\ \emph {et~al.}(2016)\citenamefont {Maurand} \emph {et~al.}}]{Maurand}%
  \BibitemOpen
  \bibfield  {author} {\bibinfo {author} {\bibfnamefont {R.}~\bibnamefont {Maurand}} \emph {et~al.},\ }\href@noop {} {\bibfield  {journal} {\bibinfo  {journal} {Nat. Commun.}\ }\textbf {\bibinfo {volume} {7}},\ \bibinfo {pages} {13575} (\bibinfo {year} {2016})}\BibitemShut {NoStop}%
\bibitem [{\citenamefont {George}\ \emph {et~al.}(2025)\citenamefont {George} \emph {et~al.}}]{George}%
  \BibitemOpen
  \bibfield  {author} {\bibinfo {author} {\bibfnamefont {H.~C.}\ \bibnamefont {George}} \emph {et~al.},\ }\href@noop {} {\bibfield  {journal} {\bibinfo  {journal} {Nano Lett.}\ }\textbf {\bibinfo {volume} {25}},\ \bibinfo {pages} {793} (\bibinfo {year} {2025})}\BibitemShut {NoStop}%
\bibitem [{\citenamefont {Bluhm}\ \emph {et~al.}(2011)\citenamefont {Bluhm} \emph {et~al.}}]{Bluhm}%
  \BibitemOpen
  \bibfield  {author} {\bibinfo {author} {\bibfnamefont {H.}~\bibnamefont {Bluhm}} \emph {et~al.},\ }\href@noop {} {\bibfield  {journal} {\bibinfo  {journal} {Nat. Phys.}\ }\textbf {\bibinfo {volume} {7}},\ \bibinfo {pages} {109} (\bibinfo {year} {2011})}\BibitemShut {NoStop}%
\bibitem [{\citenamefont {Veldhorst}\ \emph {et~al.}(2014)\citenamefont {Veldhorst} \emph {et~al.}}]{Veldhorst}%
  \BibitemOpen
  \bibfield  {author} {\bibinfo {author} {\bibfnamefont {M.}~\bibnamefont {Veldhorst}} \emph {et~al.},\ }\href@noop {} {\bibfield  {journal} {\bibinfo  {journal} {Nat. Nanotechnol.}\ }\textbf {\bibinfo {volume} {9}},\ \bibinfo {pages} {981} (\bibinfo {year} {2014})}\BibitemShut {NoStop}%
\bibitem [{\citenamefont {Kobayashi}\ \emph {et~al.}(2021)\citenamefont {Kobayashi} \emph {et~al.}}]{Kobayashi}%
  \BibitemOpen
  \bibfield  {author} {\bibinfo {author} {\bibfnamefont {T.}~\bibnamefont {Kobayashi}} \emph {et~al.},\ }\href@noop {} {\bibfield  {journal} {\bibinfo  {journal} {Nat. Mater.}\ }\textbf {\bibinfo {volume} {20}},\ \bibinfo {pages} {38} (\bibinfo {year} {2021})}\BibitemShut {NoStop}%
\bibitem [{\citenamefont {Mills}\ \emph {et~al.}(2022)\citenamefont {Mills} \emph {et~al.}}]{Mills}%
  \BibitemOpen
  \bibfield  {author} {\bibinfo {author} {\bibfnamefont {A.~R.}\ \bibnamefont {Mills}} \emph {et~al.},\ }\href@noop {} {\bibfield  {journal} {\bibinfo  {journal} {Sci. Adv.}\ }\textbf {\bibinfo {volume} {8}},\ \bibinfo {pages} {eabn5130} (\bibinfo {year} {2022})}\BibitemShut {NoStop}%
\bibitem [{\citenamefont {Hendrickx}\ \emph {et~al.}(2020)\citenamefont {Hendrickx} \emph {et~al.}}]{Hendrickx}%
  \BibitemOpen
  \bibfield  {author} {\bibinfo {author} {\bibfnamefont {N.~W.}\ \bibnamefont {Hendrickx}} \emph {et~al.},\ }\href@noop {} {\bibfield  {journal} {\bibinfo  {journal} {Nat. Commun.}\ }\textbf {\bibinfo {volume} {11}},\ \bibinfo {pages} {3478} (\bibinfo {year} {2020})}\BibitemShut {NoStop}%
\bibitem [{\citenamefont {Xue}\ \emph {et~al.}(2022)\citenamefont {Xue} \emph {et~al.}}]{Xue}%
  \BibitemOpen
  \bibfield  {author} {\bibinfo {author} {\bibfnamefont {X.}~\bibnamefont {Xue}} \emph {et~al.},\ }\href@noop {} {\bibfield  {journal} {\bibinfo  {journal} {Nature}\ }\textbf {\bibinfo {volume} {601}},\ \bibinfo {pages} {343} (\bibinfo {year} {2022})}\BibitemShut {NoStop}%
\bibitem [{\citenamefont {Wecker}\ \emph {et~al.}(2014)\citenamefont {Wecker} \emph {et~al.}}]{Wecker}%
  \BibitemOpen
  \bibfield  {author} {\bibinfo {author} {\bibfnamefont {D.}~\bibnamefont {Wecker}} \emph {et~al.},\ }\href@noop {} {\bibfield  {journal} {\bibinfo  {journal} {Phys. Rev. A}\ }\textbf {\bibinfo {volume} {90}},\ \bibinfo {pages} {022305} (\bibinfo {year} {2014})}\BibitemShut {NoStop}%
\bibitem [{\citenamefont {de~Leon}\ \emph {et~al.}(2021)\citenamefont {de~Leon} \emph {et~al.}}]{deLeon}%
  \BibitemOpen
  \bibfield  {author} {\bibinfo {author} {\bibfnamefont {N.~P.}\ \bibnamefont {de~Leon}} \emph {et~al.},\ }\href@noop {} {\bibfield  {journal} {\bibinfo  {journal} {Science}\ }\textbf {\bibinfo {volume} {372}},\ \bibinfo {pages} {eabb2823} (\bibinfo {year} {2021})}\BibitemShut {NoStop}%
\bibitem [{\citenamefont {Dzundza}\ \emph {et~al.}(2020)\citenamefont {Dzundza} \emph {et~al.}}]{Dzundza}%
  \BibitemOpen
  \bibfield  {author} {\bibinfo {author} {\bibfnamefont {B.}~\bibnamefont {Dzundza}} \emph {et~al.},\ }\href@noop {} {\bibfield  {journal} {\bibinfo  {journal} {Phys. B: Condens. Matter}\ }\textbf {\bibinfo {volume} {588}},\ \bibinfo {pages} {412178} (\bibinfo {year} {2020})}\BibitemShut {NoStop}%
\bibitem [{\citenamefont {Peres}\ \emph {et~al.}(2014)\citenamefont {Peres} \emph {et~al.}}]{Peres}%
  \BibitemOpen
  \bibfield  {author} {\bibinfo {author} {\bibfnamefont {M.~L.}\ \bibnamefont {Peres}} \emph {et~al.},\ }\href@noop {} {\bibfield  {journal} {\bibinfo  {journal} {J. Appl. Phys.}\ }\textbf {\bibinfo {volume} {115}},\ \bibinfo {pages} {093704} (\bibinfo {year} {2014})}\BibitemShut {NoStop}%
\bibitem [{\citenamefont {Gomanko}\ \emph {et~al.}(2022)\citenamefont {Gomanko} \emph {et~al.}}]{Gomanko}%
  \BibitemOpen
  \bibfield  {author} {\bibinfo {author} {\bibfnamefont {M.}~\bibnamefont {Gomanko}} \emph {et~al.},\ }\href@noop {} {\bibfield  {journal} {\bibinfo  {journal} {SciPost Phys.}\ }\textbf {\bibinfo {volume} {13}},\ \bibinfo {pages} {089} (\bibinfo {year} {2022})}\BibitemShut {NoStop}%
\bibitem [{\citenamefont {ten Kate}\ \emph {et~al.}(2022)\citenamefont {ten Kate} \emph {et~al.}}]{tenKate}%
  \BibitemOpen
  \bibfield  {author} {\bibinfo {author} {\bibfnamefont {S.~C.}\ \bibnamefont {ten Kate}} \emph {et~al.},\ }\href@noop {} {\bibfield  {journal} {\bibinfo  {journal} {Nano Lett.}\ }\textbf {\bibinfo {volume} {22}},\ \bibinfo {pages} {7049} (\bibinfo {year} {2022})}\BibitemShut {NoStop}%
\bibitem [{\citenamefont {Golovach}\ \emph {et~al.}(2006)\citenamefont {Golovach} \emph {et~al.}}]{Golovach}%
  \BibitemOpen
  \bibfield  {author} {\bibinfo {author} {\bibfnamefont {V.~N.}\ \bibnamefont {Golovach}} \emph {et~al.},\ }\href@noop {} {\bibfield  {journal} {\bibinfo  {journal} {Phys. Rev. B}\ }\textbf {\bibinfo {volume} {74}},\ \bibinfo {pages} {165319} (\bibinfo {year} {2006})}\BibitemShut {NoStop}%
\bibitem [{\citenamefont {Tyryshkin}\ \emph {et~al.}(2011)\citenamefont {Tyryshkin} \emph {et~al.}}]{Tyryshkin}%
  \BibitemOpen
  \bibfield  {author} {\bibinfo {author} {\bibfnamefont {A.~M.}\ \bibnamefont {Tyryshkin}} \emph {et~al.},\ }\href@noop {} {\bibfield  {journal} {\bibinfo  {journal} {Nat. Mater.}\ }\textbf {\bibinfo {volume} {11}},\ \bibinfo {pages} {143} (\bibinfo {year} {2011})}\BibitemShut {NoStop}%
\bibitem [{\citenamefont {Nadj-Perge}\ \emph {et~al.}(2010)\citenamefont {Nadj-Perge} \emph {et~al.}}]{NadjPerge}%
  \BibitemOpen
  \bibfield  {author} {\bibinfo {author} {\bibfnamefont {S.}~\bibnamefont {Nadj-Perge}} \emph {et~al.},\ }\href@noop {} {\bibfield  {journal} {\bibinfo  {journal} {Nature}\ }\textbf {\bibinfo {volume} {468}},\ \bibinfo {pages} {1084} (\bibinfo {year} {2010})}\BibitemShut {NoStop}%
\bibitem [{\citenamefont {Kloeffel}\ \emph {et~al.}(2013)\citenamefont {Kloeffel} \emph {et~al.}}]{Kloeffel}%
  \BibitemOpen
  \bibfield  {author} {\bibinfo {author} {\bibfnamefont {C.}~\bibnamefont {Kloeffel}} \emph {et~al.},\ }\href@noop {} {\bibfield  {journal} {\bibinfo  {journal} {Phys. Rev. B}\ }\textbf {\bibinfo {volume} {88}},\ \bibinfo {pages} {241405(R)} (\bibinfo {year} {2013})}\BibitemShut {NoStop}%
\bibitem [{\citenamefont {Huang}\ and\ \citenamefont {Hu}(2014)}]{Huang}%
  \BibitemOpen
  \bibfield  {author} {\bibinfo {author} {\bibfnamefont {P.}~\bibnamefont {Huang}}\ and\ \bibinfo {author} {\bibfnamefont {X.}~\bibnamefont {Hu}},\ }\href@noop {} {\bibfield  {journal} {\bibinfo  {journal} {Phys. Rev. B}\ }\textbf {\bibinfo {volume} {89}},\ \bibinfo {pages} {195302} (\bibinfo {year} {2014})}\BibitemShut {NoStop}%
\bibitem [{\citenamefont {Springholz}\ and\ \citenamefont {Bauer}(2014)}]{Springholz}%
  \BibitemOpen
  \bibfield  {author} {\bibinfo {author} {\bibfnamefont {G.}~\bibnamefont {Springholz}}\ and\ \bibinfo {author} {\bibfnamefont {G.}~\bibnamefont {Bauer}},\ }\href@noop {} {\emph {\bibinfo {title} {Semiconductors, IV-VI}}}\ (\bibinfo  {publisher} {John Wiley {\&} Sons, Hobokeh, New Jersey, US},\ \bibinfo {year} {2014})\BibitemShut {NoStop}%
\bibitem [{\citenamefont {Cheng}\ \emph {et~al.}(2015)\citenamefont {Cheng} \emph {et~al.}}]{Cheng_Levy}%
  \BibitemOpen
  \bibfield  {author} {\bibinfo {author} {\bibfnamefont {G.}~\bibnamefont {Cheng}} \emph {et~al.},\ }\href@noop {} {\bibfield  {journal} {\bibinfo  {journal} {Nature}\ }\textbf {\bibinfo {volume} {521}},\ \bibinfo {pages} {196} (\bibinfo {year} {2015})}\BibitemShut {NoStop}%
\bibitem [{\citenamefont {Burkard}\ \emph {et~al.}(2023)\citenamefont {Burkard} \emph {et~al.}}]{Burkard}%
  \BibitemOpen
  \bibfield  {author} {\bibinfo {author} {\bibfnamefont {G.}~\bibnamefont {Burkard}} \emph {et~al.},\ }\href@noop {} {\bibfield  {journal} {\bibinfo  {journal} {Rev. Mod. Phys.}\ }\textbf {\bibinfo {volume} {95}},\ \bibinfo {pages} {025003} (\bibinfo {year} {2023})}\BibitemShut {NoStop}%
\bibitem [{\citenamefont {Nichol}\ \emph {et~al.}(2017)\citenamefont {Nichol} \emph {et~al.}}]{Nichol}%
  \BibitemOpen
  \bibfield  {author} {\bibinfo {author} {\bibfnamefont {J.}~\bibnamefont {Nichol}} \emph {et~al.},\ }\href@noop {} {\bibfield  {journal} {\bibinfo  {journal} {npj Quantum Inf.}\ }\textbf {\bibinfo {volume} {3}},\ \bibinfo {pages} {3} (\bibinfo {year} {2017})}\BibitemShut {NoStop}%
\bibitem [{\citenamefont {Zhang}\ \emph {et~al.}(2024)\citenamefont {Zhang} \emph {et~al.}}]{ZhangTongxie}%
  \BibitemOpen
  \bibfield  {author} {\bibinfo {author} {\bibfnamefont {T.~X.}\ \bibnamefont {Zhang}} \emph {et~al.},\ }\href@noop {} {\bibfield  {journal} {\bibinfo  {journal} {ACS Appl. Mater. Interfaces}\ }\textbf {\bibinfo {volume} {16}},\ \bibinfo {pages} {54837} (\bibinfo {year} {2024})}\BibitemShut {NoStop}%
\bibitem [{\citenamefont {Nagarajan}\ \emph {et~al.}(2024)\citenamefont {Nagarajan} \emph {et~al.}}]{Nagarajan}%
  \BibitemOpen
  \bibfield  {author} {\bibinfo {author} {\bibfnamefont {S.}~\bibnamefont {Nagarajan}} \emph {et~al.},\ }\href@noop {} {\bibfield  {journal} {\bibinfo  {journal} {Adv. Mater. Interfaces}\ }\textbf {\bibinfo {volume} {11}},\ \bibinfo {pages} {2300600} (\bibinfo {year} {2024})}\BibitemShut {NoStop}%
\bibitem [{\citenamefont {van~der Wiel}\ \emph {et~al.}(2002)\citenamefont {van~der Wiel} \emph {et~al.}}]{VanDerWiel}%
  \BibitemOpen
  \bibfield  {author} {\bibinfo {author} {\bibfnamefont {W.~G.}\ \bibnamefont {van~der Wiel}} \emph {et~al.},\ }\href@noop {} {\bibfield  {journal} {\bibinfo  {journal} {Rev. Mod. Phys.}\ }\textbf {\bibinfo {volume} {75}},\ \bibinfo {pages} {1} (\bibinfo {year} {2002})}\BibitemShut {NoStop}%
\bibitem [{\citenamefont {Hanson}\ \emph {et~al.}(2007)\citenamefont {Hanson} \emph {et~al.}}]{Hanson}%
  \BibitemOpen
  \bibfield  {author} {\bibinfo {author} {\bibfnamefont {R.}~\bibnamefont {Hanson}} \emph {et~al.},\ }\href@noop {} {\bibfield  {journal} {\bibinfo  {journal} {Rev. Mod. Phys.}\ }\textbf {\bibinfo {volume} {79}},\ \bibinfo {pages} {1217} (\bibinfo {year} {2007})}\BibitemShut {NoStop}%
\bibitem [{\citenamefont {Byard}\ \emph {et~al.}(2025)\citenamefont {Byard} \emph {et~al.}}]{Byard}%
  \BibitemOpen
  \bibfield  {author} {\bibinfo {author} {\bibfnamefont {S.}~\bibnamefont {Byard}} \emph {et~al.},\ }\href@noop {} {\bibfield  {journal} {\bibinfo  {journal} {arXiv:2504.05221}\ } (\bibinfo {year} {2025})}\BibitemShut {NoStop}%
\bibitem [{\citenamefont {Sapmaz}\ \emph {et~al.}(2005)\citenamefont {Sapmaz} \emph {et~al.}}]{Sapmaz}%
  \BibitemOpen
  \bibfield  {author} {\bibinfo {author} {\bibfnamefont {S.}~\bibnamefont {Sapmaz}} \emph {et~al.},\ }\href@noop {} {\bibfield  {journal} {\bibinfo  {journal} {Phys. Rev. B}\ }\textbf {\bibinfo {volume} {71}},\ \bibinfo {pages} {153402} (\bibinfo {year} {2005})}\BibitemShut {NoStop}%
\bibitem [{\citenamefont {Mavalankar}\ \emph {et~al.}(2016)\citenamefont {Mavalankar} \emph {et~al.}}]{Mavalankar}%
  \BibitemOpen
  \bibfield  {author} {\bibinfo {author} {\bibfnamefont {A.}~\bibnamefont {Mavalankar}} \emph {et~al.},\ }\href@noop {} {\bibfield  {journal} {\bibinfo  {journal} {Phys. Rev. B}\ }\textbf {\bibinfo {volume} {93}},\ \bibinfo {pages} {235428} (\bibinfo {year} {2016})}\BibitemShut {NoStop}%
\bibitem [{\citenamefont {Hartke}\ \emph {et~al.}(2018)\citenamefont {Hartke} \emph {et~al.}}]{Hartke}%
  \BibitemOpen
  \bibfield  {author} {\bibinfo {author} {\bibfnamefont {T.~R.}\ \bibnamefont {Hartke}} \emph {et~al.},\ }\href@noop {} {\bibfield  {journal} {\bibinfo  {journal} {PRL}\ }\textbf {\bibinfo {volume} {120}},\ \bibinfo {pages} {097701} (\bibinfo {year} {2018})}\BibitemShut {NoStop}%
\bibitem [{\citenamefont {Ono}\ \emph {et~al.}(2002)\citenamefont {Ono} \emph {et~al.}}]{Ono}%
  \BibitemOpen
  \bibfield  {author} {\bibinfo {author} {\bibfnamefont {K.}~\bibnamefont {Ono}} \emph {et~al.},\ }\href@noop {} {\bibfield  {journal} {\bibinfo  {journal} {Science}\ }\textbf {\bibinfo {volume} {297}},\ \bibinfo {pages} {1313} (\bibinfo {year} {2002})}\BibitemShut {NoStop}%
\bibitem [{\citenamefont {Johnson}\ \emph {et~al.}(2005)\citenamefont {Johnson} \emph {et~al.}}]{Johnson}%
  \BibitemOpen
  \bibfield  {author} {\bibinfo {author} {\bibfnamefont {A.~C.}\ \bibnamefont {Johnson}} \emph {et~al.},\ }\href@noop {} {\bibfield  {journal} {\bibinfo  {journal} {Phys. Rev. B}\ }\textbf {\bibinfo {volume} {72}},\ \bibinfo {pages} {165308} (\bibinfo {year} {2005})}\BibitemShut {NoStop}%
\bibitem [{\citenamefont {Petta}\ \emph {et~al.}(2005)\citenamefont {Petta} \emph {et~al.}}]{Petta}%
  \BibitemOpen
  \bibfield  {author} {\bibinfo {author} {\bibfnamefont {J.~R.}\ \bibnamefont {Petta}} \emph {et~al.},\ }\href@noop {} {\bibfield  {journal} {\bibinfo  {journal} {Science}\ }\textbf {\bibinfo {volume} {309}},\ \bibinfo {pages} {2180} (\bibinfo {year} {2005})}\BibitemShut {NoStop}%
\bibitem [{\citenamefont {Koppens}\ \emph {et~al.}(2006)\citenamefont {Koppens} \emph {et~al.}}]{Koppens}%
  \BibitemOpen
  \bibfield  {author} {\bibinfo {author} {\bibfnamefont {F.~H.~L.}\ \bibnamefont {Koppens}} \emph {et~al.},\ }\href@noop {} {\bibfield  {journal} {\bibinfo  {journal} {Nature}\ }\textbf {\bibinfo {volume} {442}},\ \bibinfo {pages} {766} (\bibinfo {year} {2006})}\BibitemShut {NoStop}%
\bibitem [{\citenamefont {Elzerman}\ \emph {et~al.}(2004)\citenamefont {Elzerman} \emph {et~al.}}]{Elzerman}%
  \BibitemOpen
  \bibfield  {author} {\bibinfo {author} {\bibfnamefont {J.~M.}\ \bibnamefont {Elzerman}} \emph {et~al.},\ }\href@noop {} {\bibfield  {journal} {\bibinfo  {journal} {Nature}\ }\textbf {\bibinfo {volume} {430}},\ \bibinfo {pages} {431} (\bibinfo {year} {2004})}\BibitemShut {NoStop}%
\bibitem [{\citenamefont {Petersson}\ \emph {et~al.}(2010)\citenamefont {Petersson} \emph {et~al.}}]{Petersson}%
  \BibitemOpen
  \bibfield  {author} {\bibinfo {author} {\bibfnamefont {K.~D.}\ \bibnamefont {Petersson}} \emph {et~al.},\ }\href@noop {} {\bibfield  {journal} {\bibinfo  {journal} {Nano Lett.}\ }\textbf {\bibinfo {volume} {10}},\ \bibinfo {pages} {2789} (\bibinfo {year} {2010})}\BibitemShut {NoStop}%
\bibitem [{\citenamefont {Crippa}\ \emph {et~al.}(2019)\citenamefont {Crippa} \emph {et~al.}}]{Crippa}%
  \BibitemOpen
  \bibfield  {author} {\bibinfo {author} {\bibfnamefont {A.}~\bibnamefont {Crippa}} \emph {et~al.},\ }\href@noop {} {\bibfield  {journal} {\bibinfo  {journal} {Nat. Commun.}\ }\textbf {\bibinfo {volume} {10}},\ \bibinfo {pages} {2776} (\bibinfo {year} {2019})}\BibitemShut {NoStop}%
\end{thebibliography}%

% Supplementary
    \clearpage
    % 4 commands below are used to change figure numbering system so they are marked S1, S2, etc.
    \setcounter{figure}{0}
    \makeatletter 
    \renewcommand{\thefigure}{S\@arabic\c@figure}
    \makeatother
    \section*{Supplementary Information}
\subsection*{Detailed summary of device fabrication process}

The devices are fabricated on undoped $\mathrm{Si/SiO_2}$ substrates patterned with local bottom gate arrays (1.5/6 nm Ti/AuPd) which consist of several sets of 5 gates each, with 70 nm pitch, and 800 nm-wide pads in-between the sets. These wider pads were intended to be used as additional bottom gates, but thus far we have not utilized them except for the accidental use in the primary Device A. AuPd is selected instead of Au for the bottom gates due to its smaller grain size. A thin layer of gate dielectric (12 nm HfO\textsubscript{2}) is deposited locally on the gate arrays, with patterning done such that the gate contacts on the side of each array are left uncovered. 

Nanowires with typical dimensions of length 5--10 $\mathrm{\upmu m}$ and width 100--200 nm are selected and transferred from the growth substrate onto the gate arrays via a fine-tipped probe that is manually controlled with a micromanipulator. An optical microscope is used to monitor the transfer process. We attempt to position the nanowires such that they lie at a $90^\circ$ angle to the bottom gates, and accordingly lie parallel with other nanowires on the same chip. This is necessary for the angled contact deposition process. Source/drain electrical contacts are deposited on the transferred nanowires such that a section of each nanowire is left open over a set of 5 bottom gates. Electrical contacts to the nanowires are made in a series of 3 depositions done in succession without breaking vacuum: 5/15 nm Ti/Au is deposited directly onto the device, then 5/60 nm Ti/Au is deposited at an angle of $+45^\circ$ lateral to the nanowire, and finally another 5/60 nm Ti/Au layer is deposited at an angle of $-45^\circ$ lateral to the nanowire. After this, the device is covered by another thin layer of gate dielectric (10 nm HfO\textsubscript{2}). Finally, a large top gate (10/180 nm Ti/Au) is added via a standard top-down deposition.

All features are patterned with electron beam lithography in PMMA. Gates and source/drain contacts are made via physical vapor deposition with electron beam evaporation. Argon ion milling is done in situ before top gates or source/drain contacts are deposited. Gate dielectric HfO\textsubscript{2} is added with atomic layer deposition. 
\par
        \newpage
        \subsection*{Additional figures}
\begin{figure}[H]
    \includegraphics[width=\columnwidth]{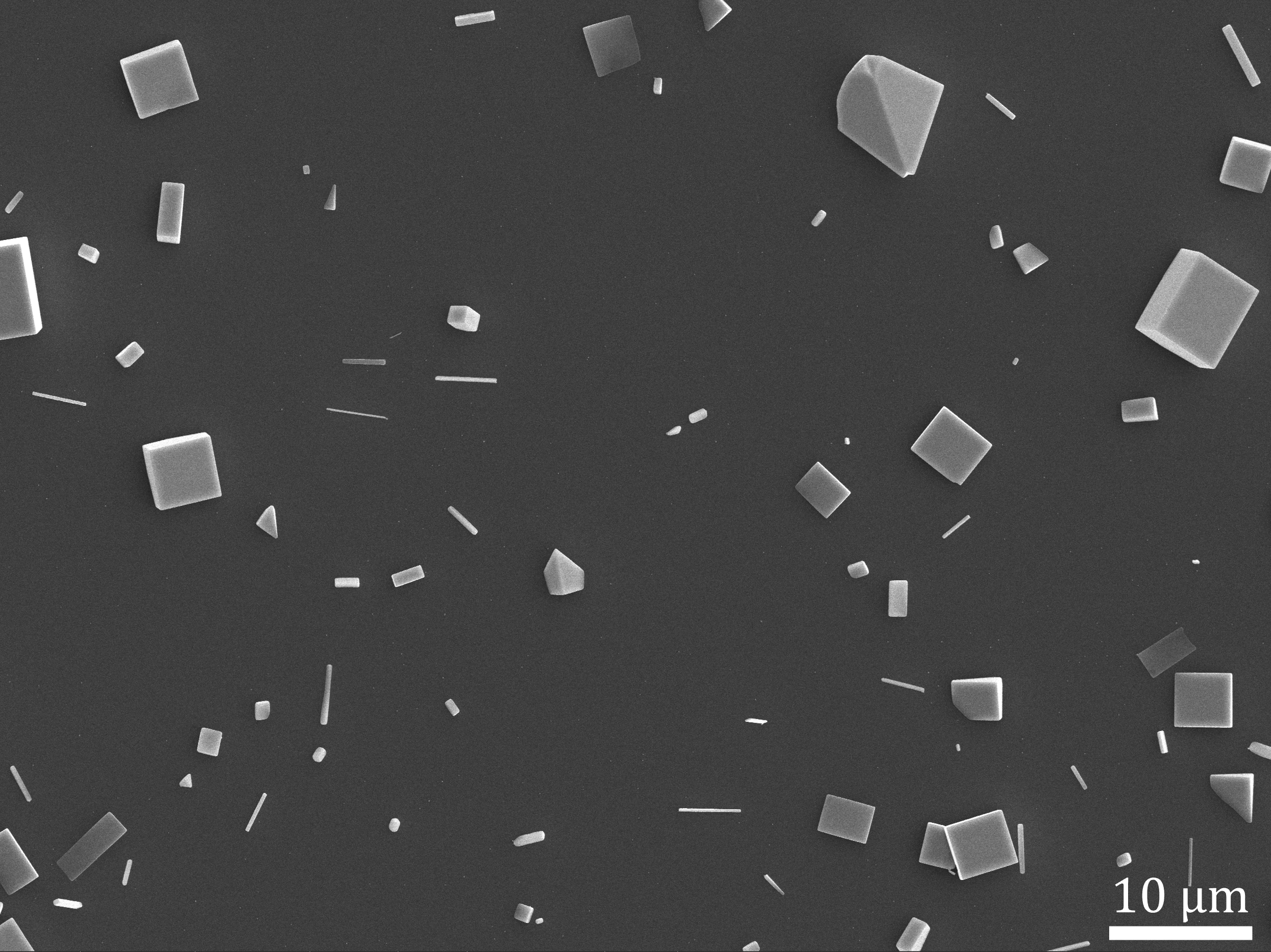}
    \centering
    \caption{SEM image of a representative section of a nanowire growth substrate. There is significant variation in the nanowires' dimensions, and an abundance of larger crystals, but nanowires of desirable length and width are still sufficiently common.}
    \label{fig:supp_growthchip}
\end{figure}
\begin{figure}[H]
    \includegraphics[width=\columnwidth]{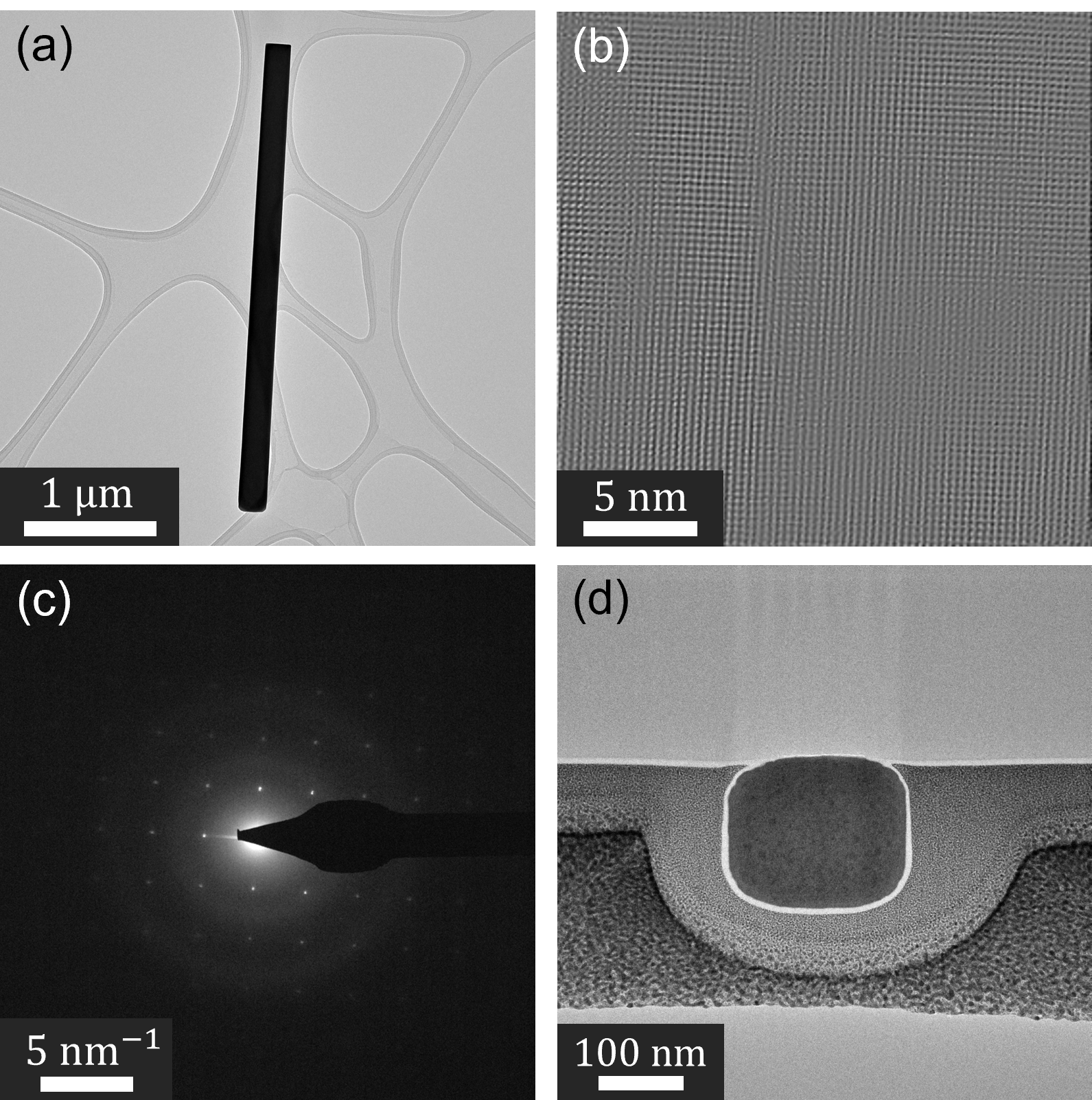}
    \centering
    \caption{TEM data for PbTe nanowires. (a) Edge-on image of nanowire. (b) FFT-filtered HR-TEM image of nanowire shown in (a). The cubic crystal structure is clear, and appears to be defect-free. (c) SAED diffraction pattern of the nanowire shown in (a), again revealing a high quality crystallinity. (d) TEM image of a cross-sectional slice of a nanowire. The white layer wrapped around the slice's edge is a carbon protection layer. The additional surrounding material is Pt deposited to protect the sample during the cutting process. The nanowire was transferred onto a silicon chip with native oxide.}
    \label{fig:supp_tem}
\end{figure}

\begin{figure}[H]
    \includegraphics[width=\columnwidth]{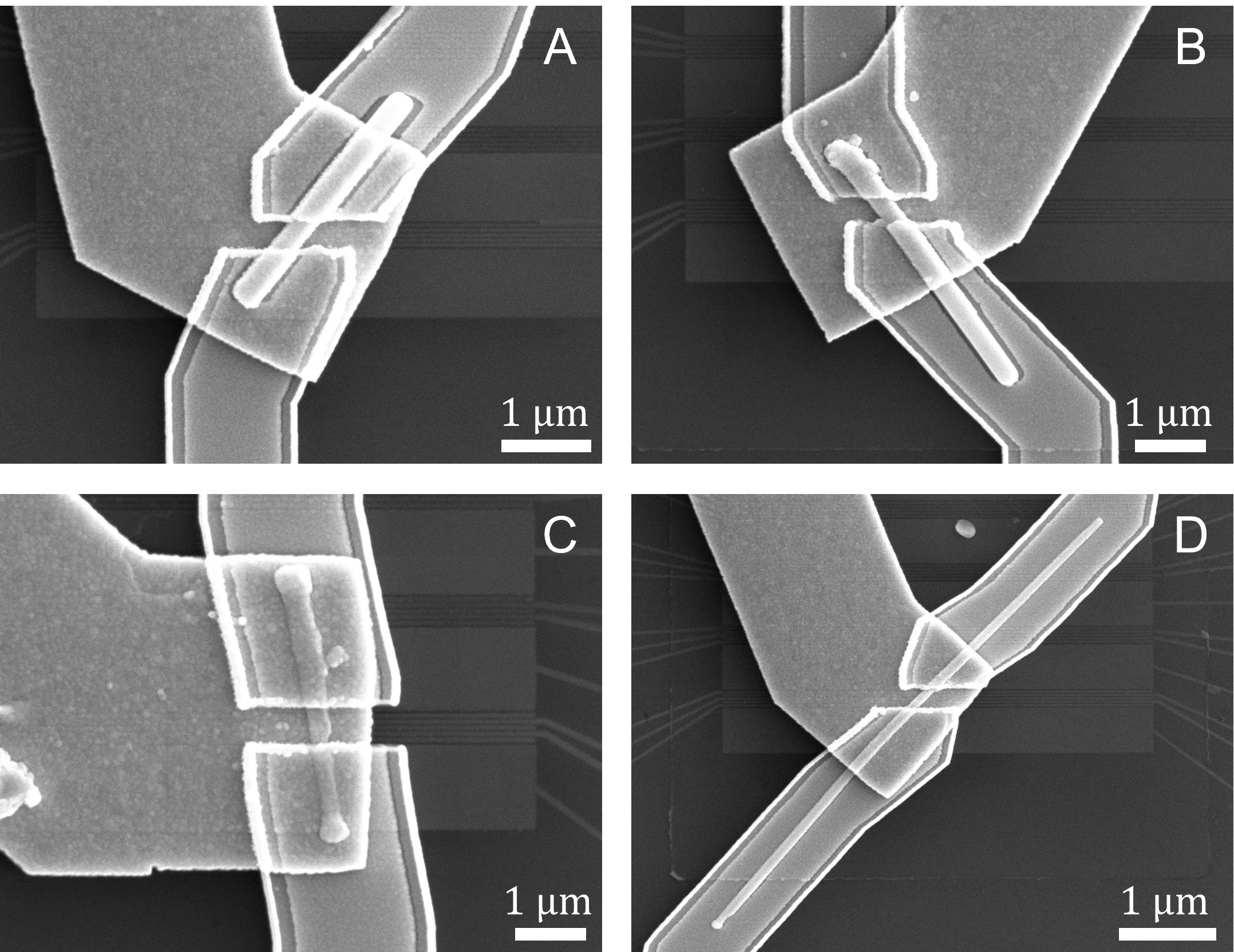}
    \centering
    \caption{SEM images of the 4 main devices (Devices A, B, C and D) measured in this study. Device A is the primary sample discussed in this work. All 4 devices were prepared on the same chip simultaneously.}
    \label{fig:supp_more_devices}
\end{figure}
\begin{figure}[H]
    \includegraphics[scale=0.6]{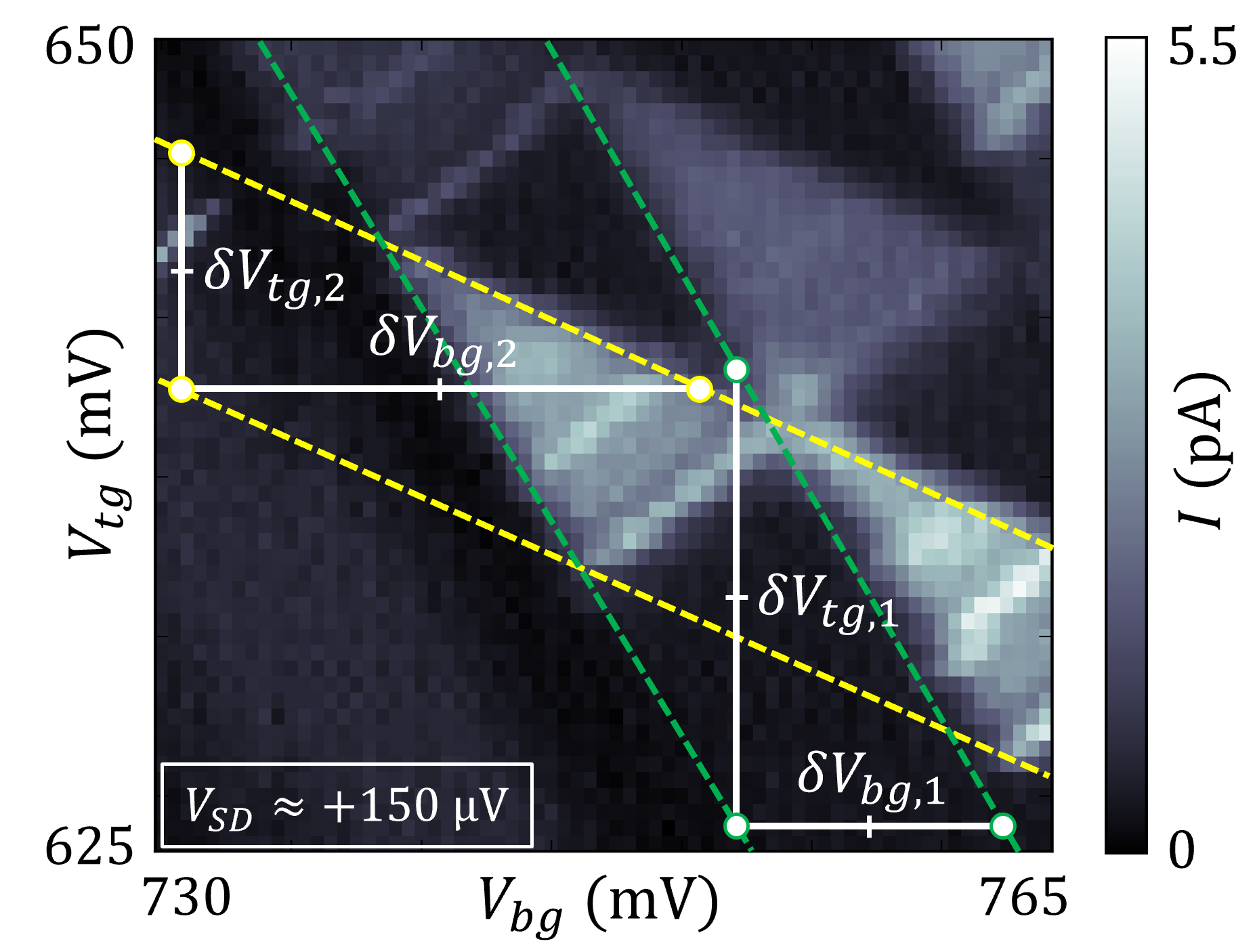}
    \centering
    \begin{center}
    \setlength{\arrayrulewidth}{0.1mm}
    \setlength{\tabcolsep}{5pt}
    \renewcommand{\arraystretch}{1.2}
    \begin{tabular}{|c|c|c|c|c|}
    \hline 
     & $\delta V_{tg,n}$ & $\alpha_{tg,n}/e$ & $\delta V_{bg,n}$ & $\alpha_{bg,n}/e$ \\
    \hline
    $n=1$ & $15.2$ mV & $0.010$ & $11.5$ mV & $0.013$ \\
    \hline
    $n=2$ & $8.2$ mV & $0.018$ & $20.8$ mV & $0.007$ \\
    \hline
    \end{tabular}
    \end{center}
    \caption{High-bias charge stability diagram from the primary sample, Device A, with a magnetic field $B=0.8\:\mathrm{T}$ applied. Additional bottom gates ``bg3'' and ``bg4'' are set to $V_{bg3}=V_{bg4}=-250$ mV. The relevant dimensions of a single triangle are indicated in the overlaid schematic. These correspond to the parameters $\delta V_{tg,n}$ and $\delta V_{tg,n}$ ($n=$ 1 or 2). We may extract the primary gate lever arms $\alpha_{tg,n}$ and $\alpha_{bg, n}$ through the general relation $\alpha_g=|eV_{SD}/\delta V_g|$, where $e$ is the elementary charge. The lever arms are the conversion factors between gate voltage and electrochemical potential for each gate-dot pair. Extracted dimensions and the corresponding lever arms are recorded in the table. It is clear that the bottom gate is most strongly coupled to dot 1, and the top gate to dot 2.}
    \label{fig:supp_lever_arms}
\end{figure}
\begin{figure}[H]
    \includegraphics[width=\columnwidth]{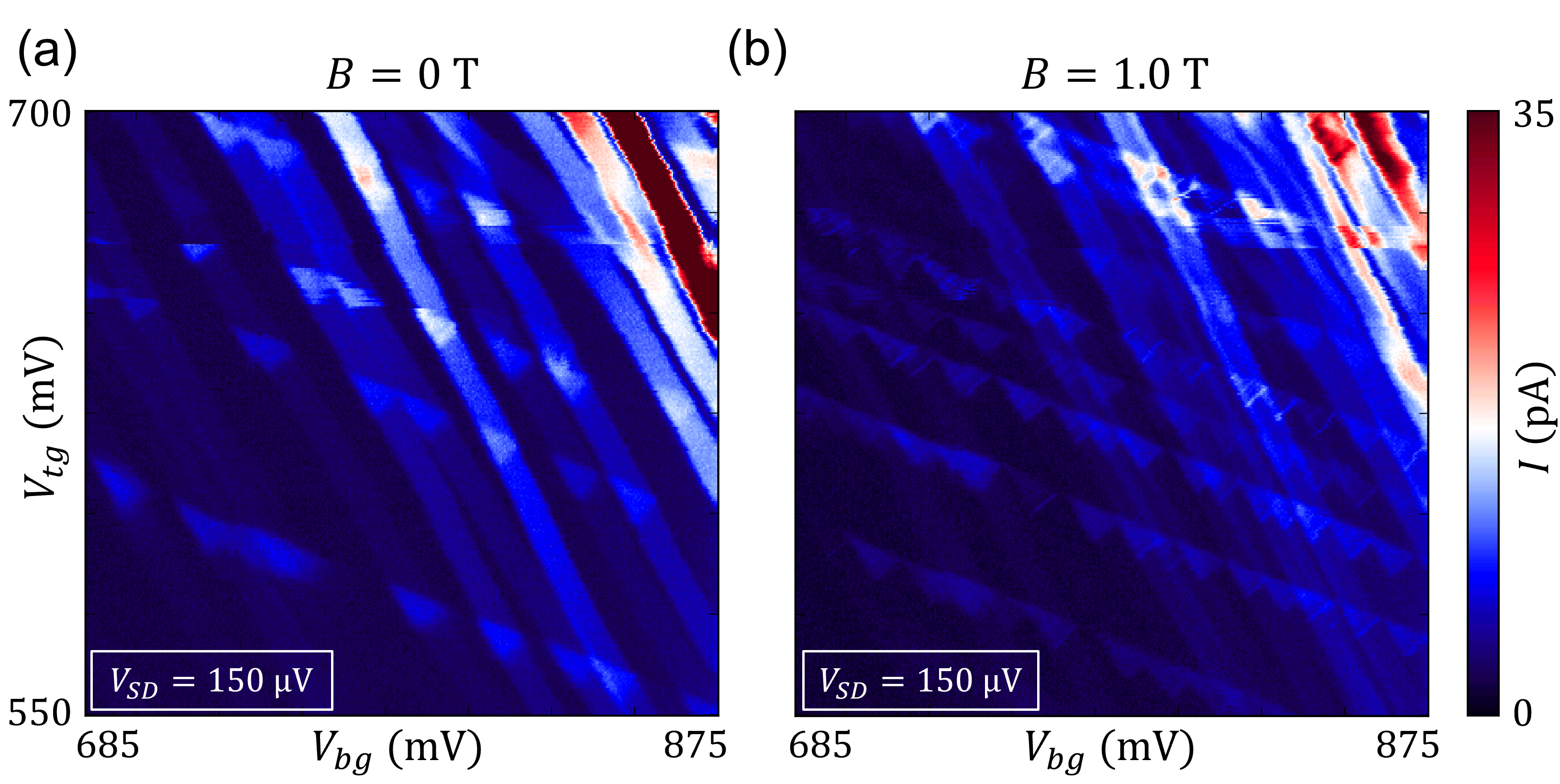}
    \centering
    \caption{Charge stability diagrams from the primary sample, Device A, at a mid-range bias both with and without a magnetic field $\vec{B}$ applied. Additional bottom gates ``bg3'' and ``bg4'' are set to $V_{bg3}=V_{bg4}=-250$~mV. When $\vec B$ is applied, all resonances split into pairs via the Zeeman effect, and thus all bias triangles split into 4, across the entire gate range used. Note that due to factors such as the close spacing of many resonances, the apparent unequal size of the dots, and the possibly unequal g-factors, the degree of splitting does not always appear consistent.}
    \label{fig:supp_extraAdata} 
\end{figure}
\begin{figure}[H]
    \includegraphics[width=\columnwidth]
    {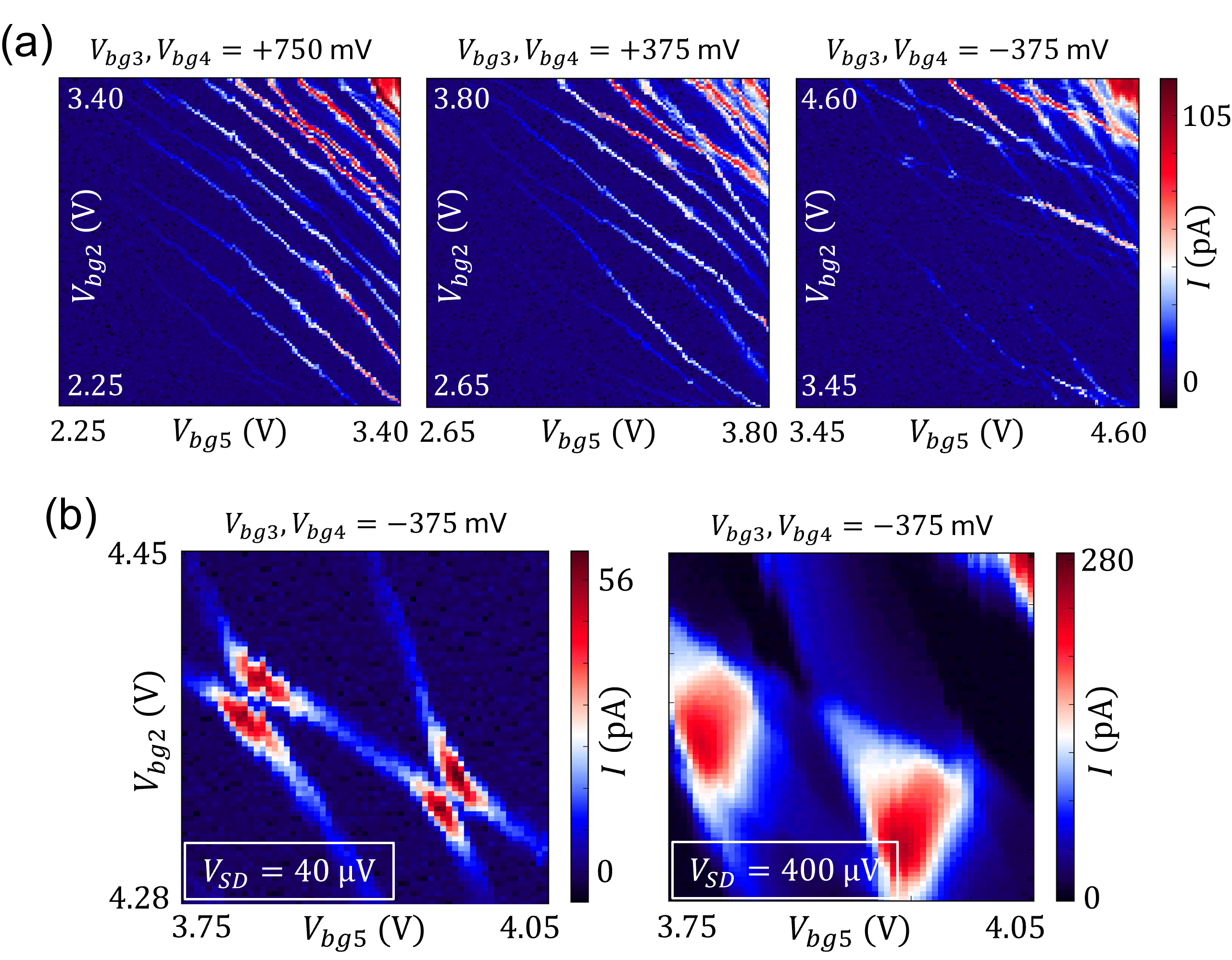}
    \centering
    \caption{Select data from device B. No magnetic field is applied. (a) Evolution of a low-bias charge stability diagram with bias $V_{SD}=20\;\mathrm{\upmu V}$ as bottom gate voltages are finely tuned. The system transitions between different transport regimes as the interdot barrier height, associated with gates ``bg3'' and ``bg4,'' is varied. This demonstrates that quantum dots in PbTe can be defined and strongly influenced by electrostatic gating. The system evidently shifts from a regime dominated by single-dot transport to one of multi-dot transport, although the multi-dot data appear to represent a triple or even higher-order dot system rather than a pure DQD. (b) Focused scans of select resonance intersections from the rightmost scan in (a). This system has negligibly small charging energy $E_C$ as usual. The offset between resonances before and after an intersection indicates a small but non-negligible mutual capacitance $C_m$ between dots. The clear avoided crossings further indicate tunnel coupling between dots. The presence of 3 or more prominent dots in this system makes it difficult to study such interactions further.}
    \label{fig:supp_Bdata}
\end{figure}
\begin{figure}[H]
    \includegraphics[width=0.99\columnwidth]{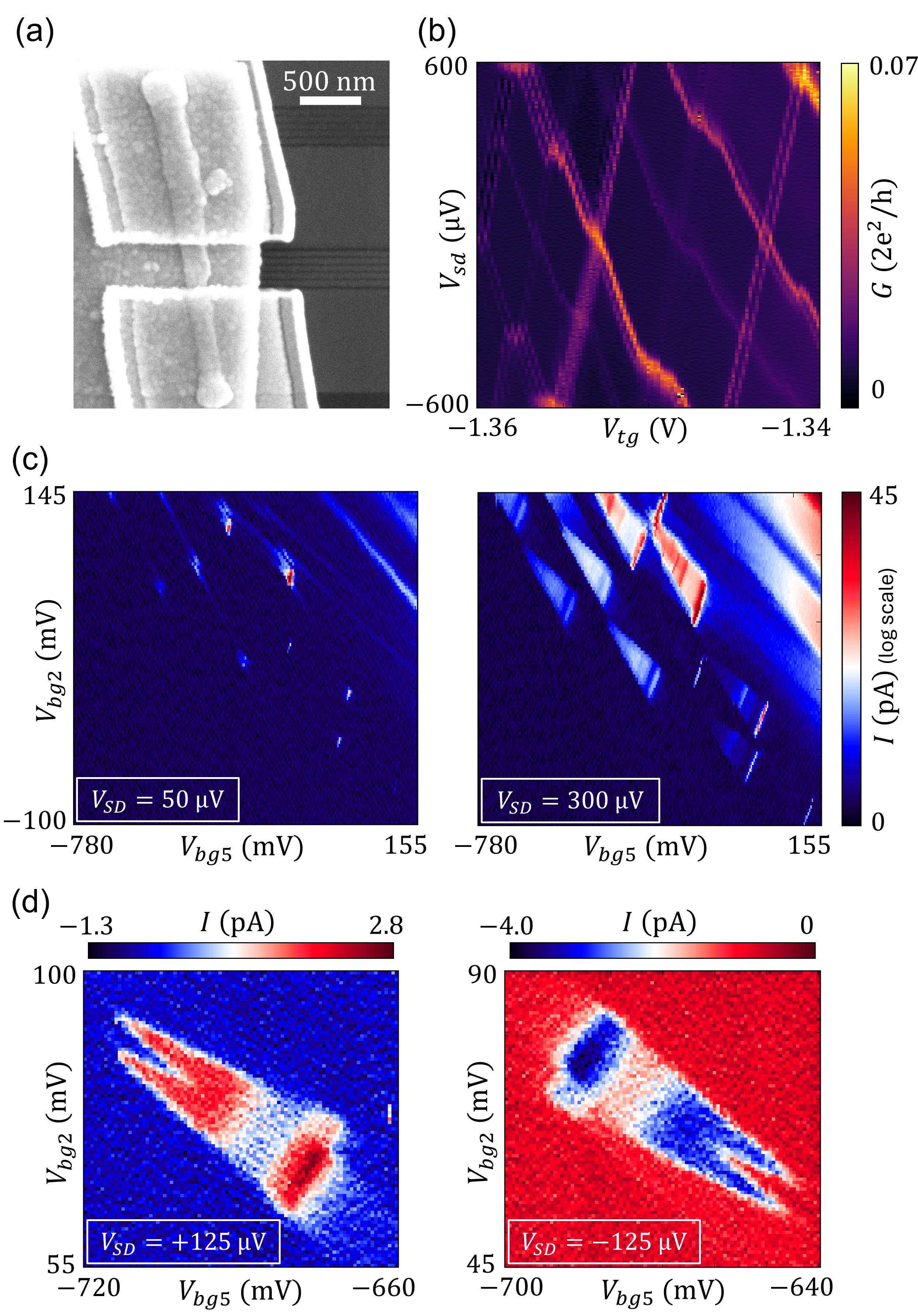}
    \centering
    \caption{Select data from device C. Unlike our typical PbTe nanowire devices, device C has finite charging energy $E_C$ and interdot mutual capacitance $C_m$. (a) SEM image of Device C. The nanowire appears to be malformed, including near where quantum dots are defined, which we argue is likely associated with the atypical behavior. (b) Coulomb diamonds plotted as differential conductance $G=dI/dV_{SD}$ vs source-drain bias $V_{SD}$ and top gate voltage $V_{tg}$. No magnetic field is applied. The diamonds alternate between large and small, with the smaller diamonds sometimes being difficult to distinguish at this resolution. This is consistent with the presence of finite $E_C$. (c) Charge stability diagrams at low and high bias with a constant magnetic field of $0.5$ T applied. (d) High-bias triangles at high positive and negative bias with a constant magnetic field of $1.5$ T applied. The triangles observed in this device overlap only partially at finite bias. These triangles contain sections of suppressed current that do not substantially vary between bias polarities.}
    \label{fig:supp_Cdata} 
\end{figure}
\begin{figure}[H]
    \includegraphics[width=\columnwidth]{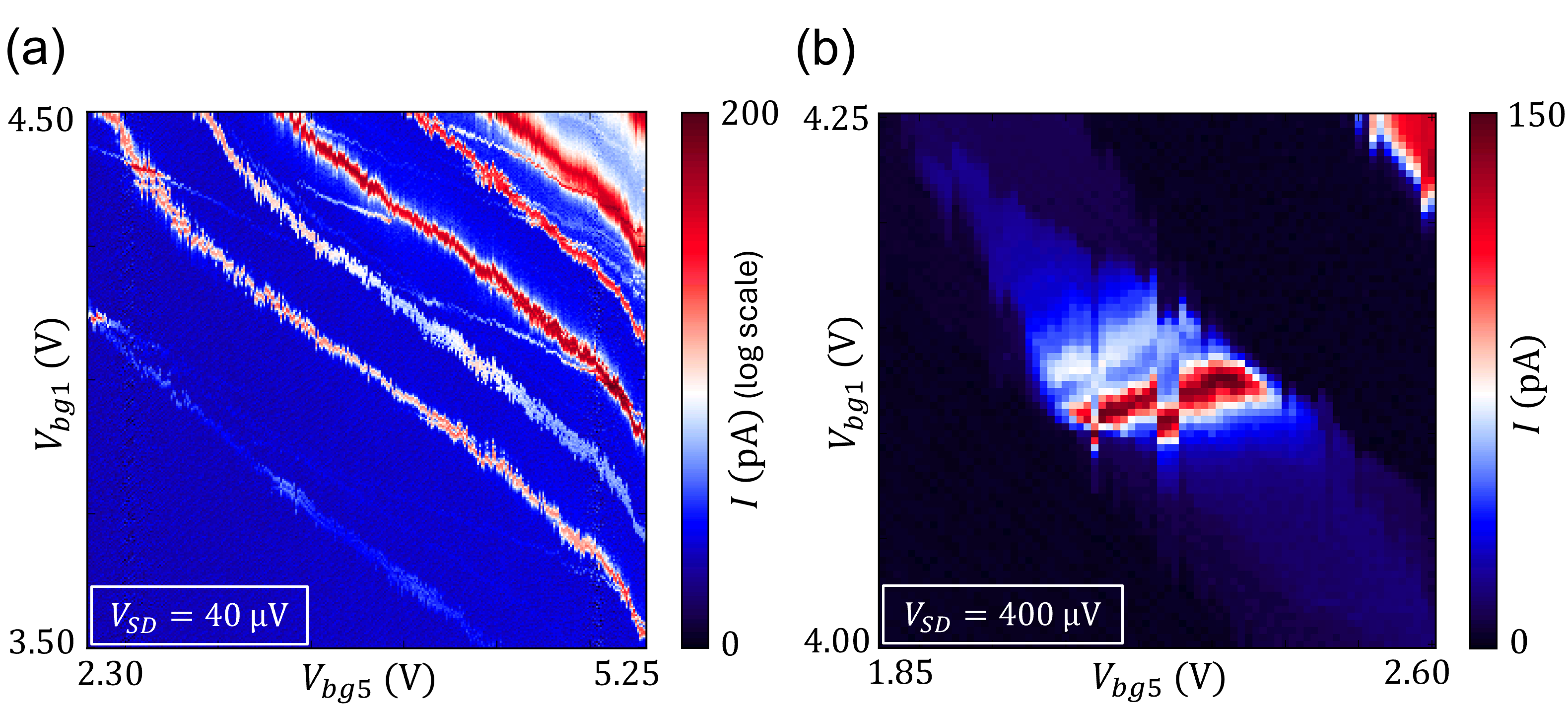}
    \centering
    \caption{Select data from device D. No magnetic field is applied. Device D is highly unstable and noisy, and thus yields relatively few useful results. (a) Low-bias charge stability diagram. It is difficult to determine the exact number and arrangement of quantum dots in this system due to irregularities in the transport resonances. (b) High-bias triangle. Despite significant noise, key features such as the triangle pair overlap and the multiple higher-energy resonances can be observed.}
    \label{fig:supp_Ddata}
\end{figure}
        
\end{document}